\documentclass[prl,aps,twocolumn,showpacs]{revtex4}
\usepackage{color}
\usepackage{array}
\usepackage{amsmath}
\usepackage{amssymb}
\usepackage{epsfig}
\usepackage{graphicx}
\usepackage{wasysym}
\usepackage{bm}

\newcommand{\be}{\begin{equation}}
\newcommand{\ee}{\end{equation}}
\newcommand{\bea}{\begin{eqnarray}}
\newcommand{\eea}{\end{eqnarray}}

\begin{document}

\title{Quantum Entanglement and Detection of Topological Order in Numerics}

\author{Tarun Grover}

\affiliation{Department of Physics, University of California,
Berkeley, CA 94720, USA}

\begin{abstract}

``Topological ordered'' phases such as gapped quantum spin-liquids and fractional quantum Hall states possess ground state degeneracy on a torus. We show that the topological nature of this degeneracy has interesting
 consequences for the entanglement structure of the degenerate ground states. This leads to a simple method for 
detecting topological order, which is ready-made for numerical schemes such as density matrix renormalization group.
The method is valid for phases that do or do not possess edge states alike. We demonstrate it by calculating the entanglement spectrum and the topological entanglement entropy for a chiral spin-liquid state on a 18 site system.
\end{abstract}

\maketitle

Entanglement properties of the ground state wavefunctions have lead to a significant progress in our numerical ability to tackle 
strongly correlated systems and have also shed light on the conceptual foundations of such systems \cite{eisert2010}. In the realm of gapped quantum phases, one of the most interesting discoveries has been the fact that the so-called ``topologically ordered phases''
\cite{wen2004, wen1989, Read89, wen1990, read1991, wen1991, sondhi, senthil2000},
that is, phases that have anyonic excitations in the bulk, 
exhibit \textit{long-range entanglement} in their ground state wave-function \cite{kitaev2003,hamma2005, kitaev2006, levin2006, chen2010}.
 One manifestation of the long range entanglement is a non-zero Topological Entanglement Entropy (TEE) $\gamma_{topo}$ \cite{hamma2005, kitaev2006, levin2006}: 
the entanglement entropy $S$ corresponding to the ground state of a two dimensional topologically ordered system for a disk-shaped region $A$ of linear length $L$ with smooth boundaries scales as 
\be
S(L) = L - \gamma_{topo} + O(1/L) \label{eq:S_L}
\ee
 The number $\gamma_{topo}$ is a robust property of the gapped phase itself and equals $\log(D)$ where $D$ is the total quantum dimension \cite{kitaev2006, levin2006}; it does not change if the 
disk-shaped region  
$A$ is deformed in any way, while maintaining its topology and keeping the boundaries smooth. This expression has been verified numerically for various topologically order states \cite{furukawa2007,haque2007,yao2010,isakov2011,frank2011, frank_smat}.
We also note that when the topological ordered state has gapless edge modes, an elegant and powerful way to identify topological order is to study the entanglement spectrum \cite{haldane} i.e. the spectrum of eigenvalues of the reduced
density matrix $\rho_A$ associated with region $A$.

 As an another application of the entanglement structure of ground state wavefunctions, the numerical methods such as density matrix renormalization group (DMRG) \cite{white92, schol} and multiscale entanglement renormalization ansatz
(MERA) \cite{vidal07} implement a real-space renormalization group that
 successively coarse-grain the system on the basis of the entanglement between a sub-region and the rest of the system. 
In particular, and what suffices to note for our purposes, DMRG involves dividing
the total system, often a cylinder or a torus, into two halves (say $A$ and $B$) that share a boundary of length $L$, and as an output, yields a Schmidt decomposed variationally optimized ground state wavefunction $|\Psi\rangle$:

\be
|\Psi\rangle = \sum_\alpha \sqrt{\lambda_\alpha}\,\,{|u_\alpha\rangle}^A \otimes {|v_\alpha\rangle}^B \label{eq:schmidt}
\ee

Here $\{\lambda_{\alpha}\}$ and $\{|u_\alpha\rangle,|v_\alpha\rangle\}$ are the Schmidt eigenvalues and eigenvectors respectively. This decomposition not only provides one 
with the correlations of various local operators, it also readily yields the Renyi entropies $S_n = -\frac{1}{n-1} \ln (\sum_\alpha {\lambda}^n_\alpha)$ as well as the von Neumann entropy 
$S_{vN} = - \sum_\alpha \lambda_\alpha \ln \lambda_\alpha$ associated with 
the bipartition. 

Juxtaposing these two developments, namely, a non-zero $\gamma_{topo}$ for topologically ordered phases and the advent of methods such as DMRG, one would think
that entanglement based numerical techniques could be used to search for new topological phases, especially when quantum Monte Carlo can not be performed because of fermionic sign problem. Indeed,
 search for quantum spin-liquids using methods that exploit entanglement is an active area of research 
\cite{capr2004, weng2006, jiang2008, even2010, yan2011}.
In particular, Ref. \cite{yan2011} has suggested that the kagome lattice antiferromagnet Heisenberg model (KAH) is a gapped quantum spin-liquid. The authors of Ref. \cite{yan2011} also found an almost
 degenerate set of two lowest lying states on a cylinder,
which is consistent with  $\mathbb{Z}_2$ spin liquid ground state. However, to establish topological order, it is desirable to extract a number such as $\gamma_{topo}$, especially because one already 
has direct access to the entanglement spectrum $\{\lambda_\alpha\}$ in DMRG. Unfortunately, there are at 
least a few practical obstructions for using DMRG to extract $\gamma_{topo}$. Eqn.\ref{eq:S_L} suggests that one may be able to extract $\gamma_{topo}$ by
 calculating $S$ as a function of system size $L$ for $L \gg\xi$, the correlation length and extrapolating 
it to $L \rightarrow 0$ to obtain $\gamma_{topo}$. In practice, this is rather difficult to implement because the two-dimensional DMRG is typically limited to cylinders or strips of width $L \lesssim 10$ 
lattice spacings 
and  the finite size effects incurred while extrapolating 
could be significant. A much more accurate method to extract TEE is to divide the total system into regions $A, B, C$ and $D$ and calculate the following \textit{tripartite} entanglement associated with regions
$A, B$ and $C$ \cite{kitaev2006, levin2006}:

\be
-\gamma_{topo} = S_A + S_B + S_C - S_{AB} - S_{BC} - S_{CA} + S_{ABC} \label{eq:tripart}
\ee

The above expression cancels out the leading `area-law' term and indeed yields an expression for $\gamma_{topo}$ that is accurate upto $\textrm{O}(1/L)$. However, calculating expression such as 
Eq.\ref{eq:tripart} within DMRG is rather challenging
and  has not been attempted so far to the best of our knowledge. Furthermore, since this method involves dividing the system into four parts (compared to two in a conventional DMRG setting) and calculating their entanglement, the finite size effects can spoil the extraction of
$\gamma_{topo}$ as they will be determined by the smallest of the regions $A$-$D$. 

\textit{In this paper, we devise a technique that can be used to cancel out the non-universal leading area law term to extract $\gamma_{topo}$, without resorting to the calculation of an expression such as Eq.\ref{eq:tripart}, 
and which can be implemented directly within DMRG or its variants}. To begin with, we note that apart from the presence of a non-zero $\gamma_{topo}$, one other important consequence of topological order is that it results in 
a degenerate set of ground states, $|\xi_i\rangle$, on a non-trivial manifold such as a torus or a cylinder ($i = 1$-$N$ where $N$ is the ground state degeneracy). The set of degenerate ground states have the property that in the thermodynamic limit, 
the expectation value of any local operator $\hat{O}$ has the form $\langle \xi_i| \hat{O}| \xi_j\rangle = O \,\delta_{ij}$, where is $O$ is a number that is independent of the index $i,j$. 
In a recent paper \cite{frank_smat} it was shown that one can extract the braiding and statistics of anyonic quasiparticles from the TEE of the degenerate set of ground states.
We now show that the generic properties of the 
entanglement spectrum of degenerate ground states impose strong constraints on the full entanglement entropy itself. This could allow one to detect topological order just using the bipartite 
entanglement in a finite sized system, provided the correlation length of local operators $\xi \ll L$, the system size, and the splitting of ground state degeneracy is negligible compared with the gap above the ground states.

We assume that the entanglement entropy of a gapped phase in two dimensions may be separated as:

\be
S = S_{local} + S_{topo} \label{eq:decomp}
\ee

Here $S_{local}$ is obtained by patching local contributions from the entangling surface  and therefore admits a curvature expansion \cite{grover2011}. On the other hand, $S_{topo}$ cannot be obtained in the 
same way, and hence is appropriately identified as the topological part of the entanglement entropy $-\gamma_{topo}$. We note that Eq.\ref{eq:decomp} is not merely a definition of $S_{topo}$ and it will be violated if, for example, there are terms of the form 
$S_{local}\times S_{topo}$. To leading order, Eq.\ref{eq:decomp} is known to hold for all gapped phases in two dimensions and this will be sufficient for our purposes.
Since the degenerate set of ground states are locally indistinguishable and possess the same correlation functions for all local operators, $S_{local}$ is same for all degenerate ground states. 
\textit{However, $S_{topological}$ in general does depend on 
the ground state, if the boundary of the region $A$ is non-contractible} \cite{dong2008, frank_smat}. In the following, we will restrict ourselves to states that only have excitations with abelian statistics.

\begin{figure}
\begin{centering}
\includegraphics[width=80mm, height = 30mm]{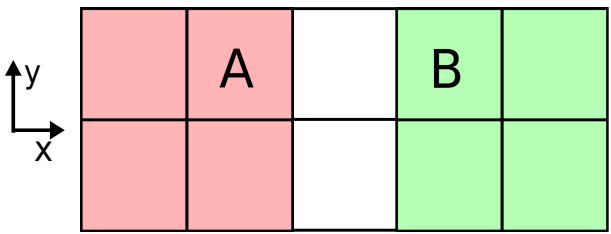}

\par\end{centering} 
\caption{Bipartition for the CSL state for which Schmidt decomposition is performed. Regions $A$ and $B$ each are of size $3 \times 3$ and the spins are located at the vertices of the 
shown lattice. The wavefunction $|0,0\rangle$ and $|0,\pi\rangle$ are obtained
by imposing periodic-periodic and periodic-antiperiodic boundary conditions respectively for the slave fermions along the $\hat{x}, \hat{y}$ directions.}
\label{fig:lat}
\end{figure}

Let us therefore consider a topologically ordered state that has $N$ degenerate ground states on a torus or cylinder. We are interested in the bipartite entanglement entropy of the ground states for an entanglement
cut similar to the one shown in Fig.\ref{fig:lat} that divides the system into two halves. As introduced in Ref.\cite{frank_smat}, there exist a special set of ground states, dubbed `minimum entropy states' (MES) 
$|\Xi_i\rangle$, $i=1$-$N$,  for which the entanglement entropy is minimum (or in other words, the magnitude of the TEE is maximum). Each MES $|\Xi_i\rangle$ is associated with a 
distinct particle type threading perpendicular to the entanglement cut.  As a concrete example, in the case of quantum Hall systems on a torus,
 $|\Xi_i\rangle$'s are eigenstates of the Wilson loop operator of the Chern-Simons gauge field that threads flux normal to the entanglement cut; while for the discrete gauge theories on a torus, $|\Xi_i\rangle$'s are the eigenstates of the 
operators that thread an electric and magnetic field normal to the cut \cite{dong2008, frank_smat}. In an abelian theory, one is free to permute the particle labels arbitrarily; therefore the Renyi TEE, and hence 
the full entanglement spectrum is same for all $|\Xi\rangle_i$. Furthermore, the same considerations imply that the Schmidt states for different MESs are orthogonal to each other:

\be
|\Xi_i\rangle = \sum_{\alpha} \sqrt{\lambda_\alpha} |u^{i}_\alpha \rangle \otimes |v^{i}_\alpha \rangle \label{eq:ee_mes}
\ee

where $\langle u^{i}_\alpha|u^{j}_\beta\rangle = \langle v^{i}_\alpha|v^{j}_\beta\rangle = \delta_{ij} \delta_{\alpha \beta}$. The factor of $\delta_{\alpha \beta}$ follows from the definition 
of Schmidt decomposition while the factor of $\delta_{ij}$ follows from the fact that each Schmidt state $|u^{i}_\alpha\rangle$ carries the distinct quantum number associated with the operator that measures the quasiparticle type 
 threading 
the entanglement cut (here, $i$). It follows from Eq.\ref{eq:ee_mes} that a generic wavefunction $|\Psi\rangle = \sum_i c_i |\Xi_i\rangle = 
\sum_{i \alpha} c_i \sqrt{\lambda_\alpha} |u^{i}_\alpha \rangle \otimes |v^{i}_\alpha \rangle$ is already Schmidt decomposed when expressed in terms of the MES \cite{footnote_schm}. 
Therefore, the $n$'th Renyi entropy corresponding to the entanglement bipartition such as that shown in Fig.\ref{fig:lat}, for the wavefunction $|\Psi\rangle$ is,

\bea
S_n & = & - \frac{\ln(\sum_i \lambda^n_\alpha)}{n-1} - \frac{\ln(\sum_i |c_i|^{2n})}{n-1} \nonumber \\
& = & S_{n.local} - b_0 \ln(D) - \frac{\ln(\sum_i |c_i|^{2n})}{n-1} \label{eq:s_ndec}
\eea

In deriving Eqn.\ref{eq:s_ndec} we have used the fact that $S_{n,local}$ is same for all degenerate ground states and that the TEE for MESs equals $b_0 \ln(D)$ where $b_0$ is the number of connected components
of the boundary of  region $A$ (for example, $b_0 = 1 \,(2)$ when the overall geometry is a cylinder (torus) \cite{degencyl}). Eqs. \ref{eq:ee_mes} and \ref{eq:s_ndec} can be used to establish topological order and extract TEE 
from bipartite entanglement entropy. We note that Eq.\ref{eq:s_ndec} agrees with the topological entanglement entropy of Chern-Simons gauge theory calculated in Ref.\cite{dong2008}.

\textit{We emphasize that the dependence of $S_n$ on the ground state wavefunction in Eq.\ref{eq:s_ndec} results solely from the fact that the boundary of region $A$ is non-contractible and therefore,
 is of topological nature}. For contractible boundaries, all topologically degenerate ground states will have the same entanglement entropy. This, along with Eq.\ref{eq:ee_mes}, provides a sharp distinction between a topologically ordered phase and a short-range entangled phase where the ground state degeneracy may result from spontaneous
 symmetry breaking \cite{symbreak}.

\begin{figure}
\begin{centering}
\includegraphics[width=90mm, height = 60mm]{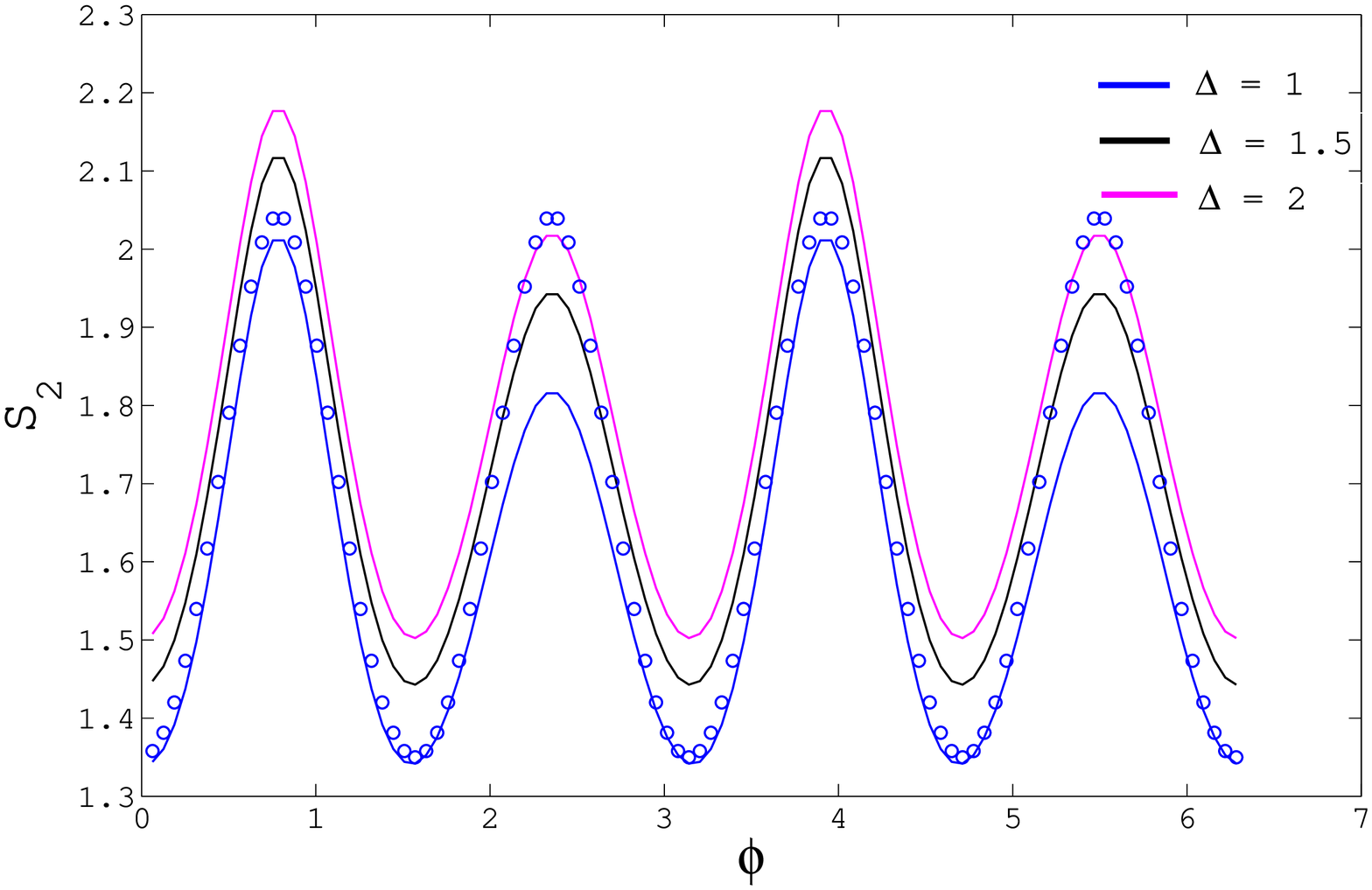}

\par\end{centering} 

\begin{centering}
\includegraphics[width=90mm, height = 70mm]{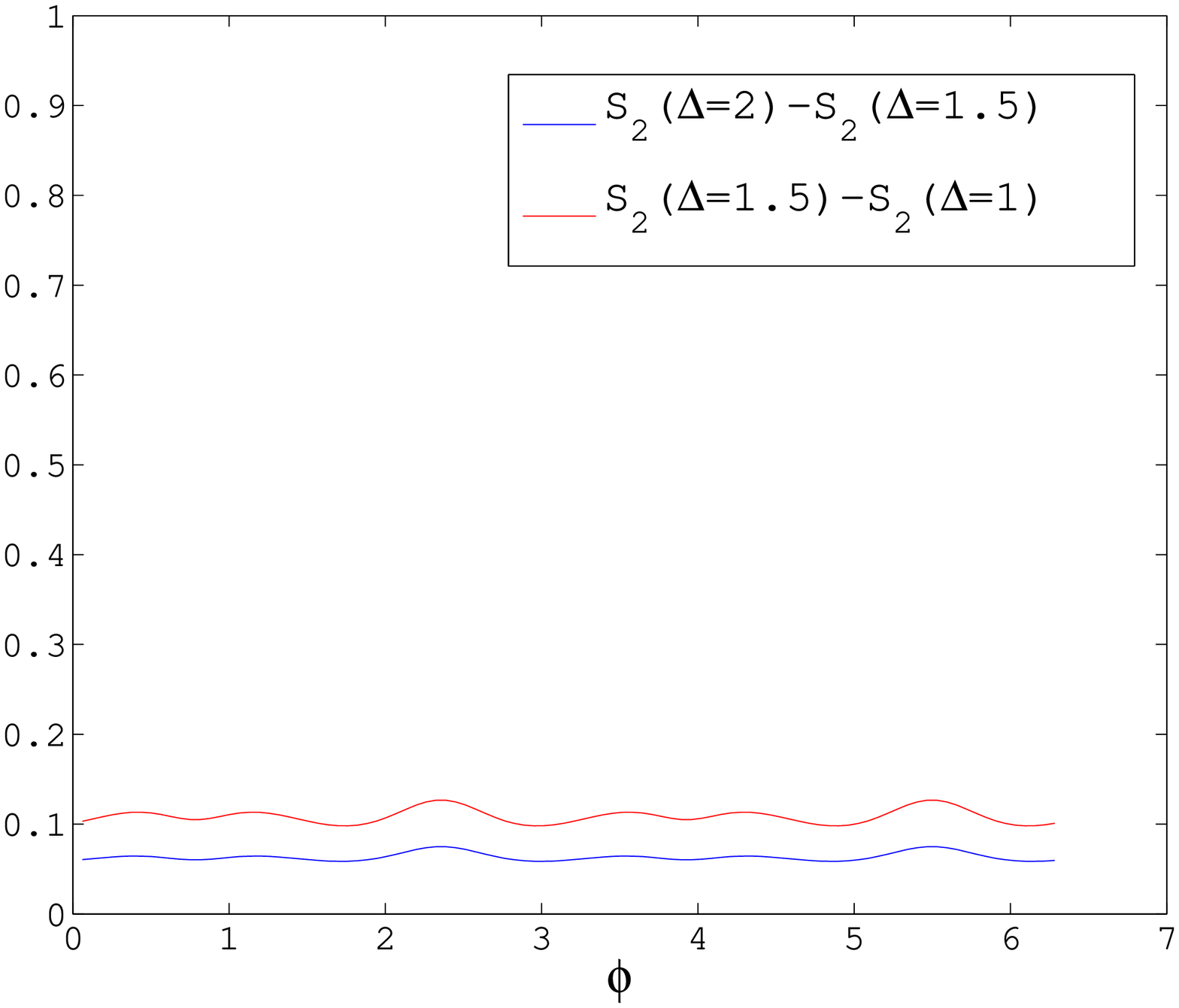}

\par\end{centering}
\caption{Above: Renyi entropy $S_2$ for the state  $|\Phi\rangle = cos(\phi) \,|0,0\rangle + sin(\phi)\, |0,\pi\rangle$ for different values of the gap $\Delta$ that enters the CSL wavefunction.
The dotted curve is the theoretical prediction in the thermodynamic limit: $S_2 = C(\Delta) - \ln(\cos^4(\phi) + \sin^4(\phi))$ where we $C(\Delta)$ is a non-universal constant (here we 
chose $C = 1.35$ so as to facilitate the comparison with the numerical data). Further, $S_{2,max}-S_{2,min} \approx 0.67$ independent of $\Delta$ which is close to the expected result of $\ln(2)$ in the thermodynamic limit.
Below: The difference between $S_2$ for two sets of two different $\Delta$'s as a function of the angle $\phi$. The difference $S_2(\Delta_1)-S_2(\Delta_2)$, is nearly $\phi$ independent in agreement with the theoretical prediction, since in
it equals $S_{2,local}(\Delta_1) - S_{2,local}(\Delta_2)$ (see Eq.\ref{eq:s_ndec}). }
\label{fig:s2phi}

\end{figure}

\begin{figure}
\begin{centering}
\includegraphics[width=90mm, height = 70mm]{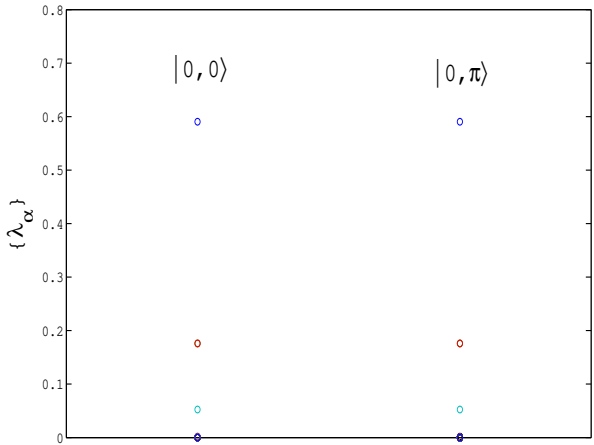}

\par\end{centering} 
\caption{Eigenvalues $\lambda_{\alpha}$ of the density matrix corresponding to the two states $|0,0\rangle$ and $|\pi,0\rangle$. Since very large number of eigenvalues have a very 
small value, only few of them are visible. The spectrum for the two states shown is identical within computer precision.}
 \label{fig:ee}
\end{figure}

As a demonstration of the ideas developed above, we first note that the Eqs.\ref{eq:ee_mes} and \ref{eq:s_ndec} are readily satisfied by fixed point wavefunctions that have zero correlation length,
 such as the ground
states of the toric code model \cite{kitaev2003} on a torus or cylinder \cite{frank_smat}. 
   As a non-trivial test, we consider the entanglement spectrum of an $SU(2)$ spin symmetric Chiral Spin Liquid (CSL)
 state \cite{kalmeyer, wenchiral, thomale} on a rectangular lattice of size $ 6 \times 3$ (Fig.\ref{fig:lat}). Note that, unlike toric code model \cite{hamma2005}, we do not have access to the 
MES apriori.
The topological order in the CSL state has already been established numerically in Ref.\cite{frank2011} by calculating $\gamma_{topo}$ and verifying that it matches the theoretical value 
$\gamma_{topo} = \log(2)/2$. The ground state wavefunction for the CSL state are obtained using a slave particle construction where one writes the spins as bilinears in 
fermions $\vec{S}=\frac12 f^\dagger_\sigma [\vec{\sigma}]_{\sigma\sigma'}f_{\sigma'}$, and assumes a chiral d-wave state for the fermions \cite{kalmeyer, wenchiral} with a mean-field BCS gap $\Delta$. To obtain
 a spin wave function, 
one Gutzwiller projects the $d_{x^2-y^2}+id_{xy}$ superconductor to one fermion per site. The degenerate ground states are obtained by putting the system on a torus 
and choosing periodic or anti-periodic boundary conditions for the slave fermions along the two torus directions \cite{Read89, wen1989, kalmeyer, wenchiral}.
We denote the ground states by the mean field fluxes in the
$\hat x$ and $\hat y$ directions as
$\left|\varphi_{1},\varphi_{2}\right\rangle$, $\varphi_{1,2} = 0, \pi$.
The two fold degeneracy of the CSL implies that only two of the four
ground states $\left|0,0\right\rangle $, $\left| \pi,0\right\rangle
$, $\left|0,\pi\right\rangle $, $\left|\pi,\pi\right\rangle $ are
linearly independent. To obtain the entanglement spectrum for a given ground state, we divide the system into two halves $A$ and $B$, each of size $3 \times 3$ and express the corresponding wavefunction as 

\be
|\Psi\rangle = \sum_{ij} \psi_{ij} |a_i\rangle \otimes |b_j\rangle
\ee

where $|a_i\rangle$ and $|b_i\rangle$ are the basis states for the region $A$ and $B$. If the regions $A$ and $B$ have $N/2$ spins each, then $\psi_{ij}$ is the $ij$'th element of a 
 $2^{N/2} \times 2^{N/2}$ matrix. The Schmidt decomposition
of the wavefunction $|\Psi\rangle$ (see Eq.\ref{eq:schmidt}) is obtained by performing the Singular Value Decomposition (SVD) of the matrix $\psi_{ij}$ numerically.

For the system size considered ($6 \times 3$ lattice), the MESs are found to correspond to the states $|\Xi_1\rangle = |0,0\rangle$ and $|\Xi_2 \rangle = |0,\pi\rangle$ (Fig.\ref{fig:s2phi}) . Because of
 finite size effects, the orthogonality of these states is not perfect 
for this system size with 
$\langle 0,\pi|0,0\rangle \approx 0.03-0.07$ for the range of mean-field gap $\Delta$ considered. Despite this, we find that the entanglement spectrum of these two states is identical (Fig.\ref{fig:ee}) in
 agreement with Eq.\ref{eq:ee_mes}. 
We do not resolve the momentum along the cut for the Schmidt states, since it is not needed for our purposes.
Further, we find that the Schmidt states $|u_\alpha^{1}\rangle,\,|u_\alpha^{2}\rangle$ corresponding to same value of the 
density-matrix eigenvalue $\lambda_\alpha$ are orthogonal in agreement with Eq.\ref{eq:ee_mes}. 
However, because of lack of the complete orthogonality of the two MES, the Schmidt states for non-identical eigenvalues are 
found to be not always orthogonal.

Fig.\ref{fig:s2phi} shows the Renyi entropy $S_2$ for the linear combination $|\Phi\rangle = cos(\phi) \,|0,0\rangle + sin(\phi)\, |0,\pi\rangle$ for different values of the mean-field d-wave BCS gap $\Delta$, that enters as a parameter
in the CSL wavefunction. We notice that the difference $S_{2,local}(\Delta_1,\phi)-S_{2,local}(\Delta_2,\phi)$ is nearly independent of $\phi$ in agreement with Eq.\ref{eq:s_ndec}: the difference depends only on $S_{local} (\Delta_1)-S_{local}(\Delta_2)$, 
thereby being only a function of $\Delta_1, \Delta_2$. Perhaps most interestingly, Eq.\ref{eq:s_ndec} predicts that for 
the CSL state, $S_{2,max} - S_{2,min} = 2 \,\gamma_{topo} = \ln(2) \approx 0.69$. We find that the numerical value of the same quantity equals $\approx 0.67$, in close agreement with the theory, and remains essentially constant
as $\Delta$ is changed. We note that though the curve $S_2(\phi)$ in Fig.\ref{fig:ee} does not agree with the theoretical prediction in the thermodynamic limit completely because of the aforementioned finite size effects
, the overall agreement it is still remarkable with a system size of only 18 sites.

Since the matrix $\psi$ is non-sparse, the numerical cost of performing SVD on a 
$2^{N/2} \times 2^{N/2}$ is nearly equal to the cost of exact diagonalizing the whole system for a local Hamiltonian. Therefore, our scheme can be implemented exactly on system sizes upto $N = 30-40$ sites for spin $S = 1/2$.  
Of course, the detection of topological order in such sizes will be contingent on whether the correlation length of various local operators $\xi \ll L$, the linear length of the system. 
As mentioned in the introduction, it is rather fortuitous that one of the best current methods of choice to solve strongly correlated systems, namely DMRG, happens to provide one with the entanglement spectrum
of the ground state wavefunction(s). Given the hints of topological order in KAH with two nearly degenerate ground states and the short correlation length of one-two unit cells observed in Ref.\cite{yan2011}, 
it may be worthwhile to apply the method described in this paper to KAH using  DMRG and exact diagonalization as well.

Finally, some the considerations developed above are also applicable to the so-called ``quantum-ordered'' phases, namely, phases with emergent gapless gauge bosons that may be coupled to bosonic/fermionic matter \cite{wen2004}. As 
an example, consider the $U(1)$ spin-liquid in three dimensions (see, e.g. Ref.\cite{hermele2004}). The low-energy spectrum of this spin-liquid contains analog of topological degenerate states whose gap scale as $\sim 1/L$ with the system
size $L$. One might think that this renders them difficult to distinguish from the excitations corresponding to fluctuations of emergent photons whose gap also scales as $1/L$. However, based on the discussion in this paper, entanglement
properties can, in fact, distinguish states that correspond to these two distinct types of excitations. In particular, for a given entanglement cut, 
using the states that are analog of topological degeneracy, one may again construct a set of minimum entropy states  that will satisfy Eq.\ref{eq:ee_mes}, while such a relation will generically not hold for the other
low-lying states.

To summarize, we described a method that can be used to detect topological order from the Schmidt decomposed degenerate ground states. The essential part consists in noticing that the  
non-universal leading term in the entanglement entropy is identical for all degenerate ground states and thus can be canceled out so as to extract the subleading universal constant, 
whose form is dictated by Eq.\ref{eq:s_ndec}, when the boundary of the bipartition is non-contractible. We also demonstrated the method by performing exact Schmidt decomposition of a topologically ordered state, the chiral spin
 liquid state.
We believe that this method can be put to practical use in numerical methods such as
DMRG and exact digonalization.

{\bf Acknowledgements:}
I thank Ari Turner, Ashvin Vishwanath, Masaki Oshikawa, Michael Levin, Michael Zaletel and Yi Zhang for very stimulating and helpful discussions.


\begin{thebibliography}{7}
\bibitem{degencyl} For Eq.\ref{eq:s_ndec} to be useful for the cylinder geometry, to begin with, 
one of course needs to make sure that the topological phase indeed admits degenerate ground states on a cylinder.
\end{thebibliography}
\end{document}